\documentclass[12pt]{article} % single column, double spaced

\usepackage{ol}

\usepackage{graphicx}

\usepackage{hyperref}
\usepackage{amsmath}

\begin{document}

%\twocolumn[ %% activate for two-column option

\title{Power dependent switching of nonlinear trapping\\by local photonic potentials}

\author{Y. Shavit,$^{1}$ Y. Linzon,$^{1,*}$ S. Bar-Ad,$^{1}$ R. Morandotti,$^{2}$ M. Volatier-Ravat,$^{3}$ V. Aimez,$^{3}$ and R. Ares$^{3}$}
\address{$^{1}$School of Physics and Astronomy, Faculty of Exact Sciences, Tel-Aviv University, Tel Aviv 69978, Israel\\
$^{2}$Universite' du Quebec, Institute National de la Recherche Scientifique, Varennes, Quebec J3X 1S2, Canada\\
$^{3}$Centre de Recherche en Nanofabrication et en Nanocaracterisation, Universite de Sherbrooke, Sherbrooke, J1K2R1, Canada\\
$^{*}$Corresponding author: linzonyo@post.tau.ac.il }

% Do not use \email or \homepage here. E-mail and URL can be given just before references.

\begin{abstract}We study experimentally and numerically the nonlinear scattering of wave packets by local multi-site guiding centers embedded in a continuous dielectric medium, as a function of the input power and angle of incidence. The extent of trapping into the linear modes of different sites is manipulated as a function of both the input power and incidence angle, demonstrating power-controlled switching of nonlinear trapping by local photonic potentials.\end{abstract}


%] %% activate for two-column option

\noindent Nonlinear two-dimensional (2D) wave scattering by local
potential discontinuities has been the subject of extensive
theoretical studies [1-9], and was recently realized experimentally
[10]. In the optical realization of 2D wave packet scattering in
planar waveguides [10], photonic scattering centers are formed by
local modulations of the refractive index, embedded in an otherwise
homogeneous single-mode planar dielectric waveguide [see Figs. 1(a)
and 1(b)]. Such centers, or "defects", can be either anti-guiding or
guiding, with the latter type supporting internal normal modes. The
normal modes can be excited by direct-excitation [Fig. 1(a)].
Radiation modes, coupled via side-excitation at an angle of
incidence $\theta$ towards the center [Fig. 1(b)], scatter from the
local structure. In the linear-wave regime, the propagation of
normal modes and the scattering of radiation modes are fully
characterized by the structure's plane wave transmission spectrum
[Figs. 1(c) and 1(d)]; Normal modes and linear radiation modes are
also decoupled, so that energy conservation of the scattered
transmitted and reflected wave packets is fulfilled: $R+T=1$.

\begin{figure}[b]
\centerline{\includegraphics[width=8cm]{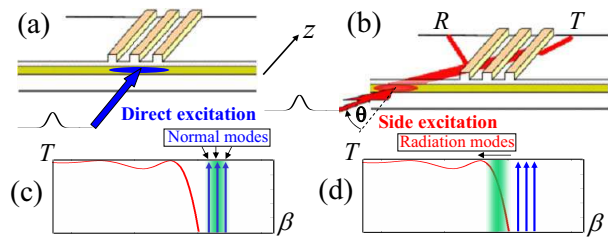}}
\caption{(Color online) (a) Directly-excited normal modes of a local
photonic structure, and (b) side-excited radiation modes scattering
from this structure; (c), (d) a characteristic plane wave
transmission spectrum. In the linear regime direct-excitations
couple to normal modes (c), while side-excitations couple to
radiation modes (d), each of which exist in different propagation
constant ($\beta$) regimes. Shaded regions in (c) and (d) indicate
spatial frequency bands of corresponding input wave packets.}
\end{figure}

In the presence of a focusing Kerr nonlinearity, the scattering
dynamics can change dramatically. In particular, the nonlinearity
can couple light between the radiation modes and normal modes
through the nonlinearity-induced change of the propagation constant
[10], provided that the local modulations are shallow and a
near-grazing angle of incidence (small $\theta$) is used. A fraction
of the incident light, having a larger effective propagation
constant, can then couple to the linear modes of the local
scatterers and trap there. In contrast with localized nonlinear
waves, which are new bound states that form in \emph{gaps} of a
corresponding linear system (such as discrete [11,12], gap [13,14]
and surface [15,16] solitons), here the nonlinear trapping (NLT) is
into pre-existing linear modes of a local guiding scattering center
(GSC). While switching of the NLT between the constituent sites of
multi-site local GSCs has been demonstrated in silica glass as a
function of $\theta$ [10], \emph{power dependent} switching has not
been previously realized. This is due to the small strengths of
available nonlinearities compared to the refractive index variations
in previous experiments.

\begin{figure}[t]
\centerline{\includegraphics[width=8cm]{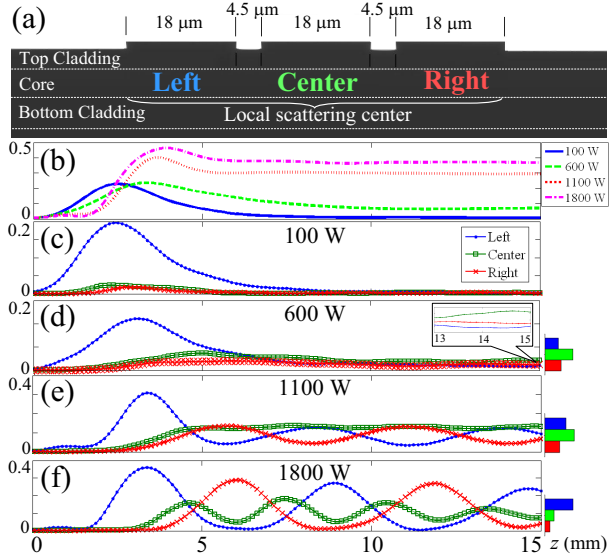}}
\caption{(Color online) (a) An illustration of the sample's
cross-section. (b)-(f) Propagation dynamics from numerical
scattering simulations with $\theta$=6 mrad; (b) fraction of the
total power trapped within the entire GSC for various input powers
in the range 100-1800 W; (c)-(f) fraction of the total power within
each site for various input powers; (c) 100 W; (d) 600 W; (e) 1100
W; (f) 1800 W. Circles, left site; squares, central site; crosses,
right site. Shown on the right are the output ($z$=15 mm) energy
distributions among the GSC sites}
\end{figure}

In this paper we report a realization of NLT switching, as a
function of the input power, between adjacent sites in a
specifically designed local GSC embedded in a nonlinear planar
AlGaAs waveguide. In the geometry considered here the Kerr
nonlinearity is comparable to the spectral spacing between normal
modes of different sites in the GSC, and the normal modes'
propagation constants are close to the radiation modes' edge [see
Figs. 1(c) and 1(d)]. A cross section of the sample is shown in Fig.
2(a). The cross section is constant along the propagation direction
$z$, and the light was coupled to the 1.5 $\mu$m-thick core layer,
which was deposited on top of a 1.5 $\mu$m-thick bottom cladding.
While the homogeneous region surrounding the local scattering center
has a top cladding layer of thickness 0.22 $\mu$m, defining an
effective refractive index of 3.32 in the transverse direction, in
the GSC sites the top layer is 1.5 $\mu$m-thick, defining an
effective index which is larger than in the homogeneous region by
0.1\%. Each of the three 18 $\mu$m-wide sites then supports 3 normal
modes. While directly excited local GSC modes are strongly-coupled
to each other, they are decoupled from radiation modes in the linear
regime.

Switching of the nonlinearly trapped power is first demonstrated in
numerical simulations of the 1+1 nonlinear Schr\"{o}dinger equation,
using the beam propagation method [17]. The input beam is a 20
$\mu$m-wide Gaussian beam, centered 60 $\mu$m away from the center
of the GSC and launched towards it at an angle of 6 mrad. Figure
2(b) shows propagation of the overall power in the GSC. For low
input powers [solid line in Fig. 2(b)] the power leaks out of the
GSC as it propagates along the $z$ direction, and practically no
power remains in the structure at the output position ($z$=15 mm,
corresponding to 3.1 diffraction lengths). This exponential decay of
power within the GSC is characteristic of linear scattering. As the
power is increased to nonlinear levels [dashed, dotted and
dash-dotted lines in Fig. 2(b)] an increasing fraction of the power
becomes trapped in the GSC region, due to NLT in the GSC's normal
modes. Figs. 2(c)-2(f) show propagation of the power fraction within
each site individually. At nonlinear powers, even though the overall
trapped power remains constant indefinitely, it oscillates between
the GSC sites with well determined amplitudes and phases, as evident
by Figs. 2(e) and 2(f). By varying the beam intensity, the nonlinear
coupling between adjacent GSC sites is detuned, thus influencing the
oscillations (both amplitude and phase) of the beam as it propagates
through the GSC. The different oscillation amplitudes and slight
phase shifts are eventually observed as different power
distributions among the GSC sites at any output position. The
alternation in the output power distribution can be conveniently
described as a shift of its' center of mass. Figure 3 shows the
output power's center of mass position as a function of the input
power. The center of mass moves initially to the right [Fig. 3(a)]
and then to the left [Figs. 3(b) and 3(c)]. Note that as the power
is further increased the center of mass shifts back to the right, in
agreement with the oscillatory nature of the trapped power dynamics.

\begin{figure}[t]
\centerline{\includegraphics[width=8cm]{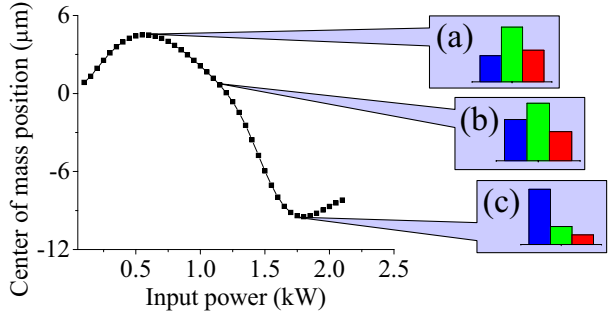}}
\caption{(Color online) Position of the center of mass at the output
($z$=15 mm) as a function of the input power, corresponding to the
numerical simulations shown in Figs. 2(b)-2(f). Insets: output power
distributions among the GSC sites at (a) 600 W; (a) 1100 W; (c) 1800
W.}
\end{figure}

\begin{figure}[t]
\centerline{\includegraphics[width=8cm]{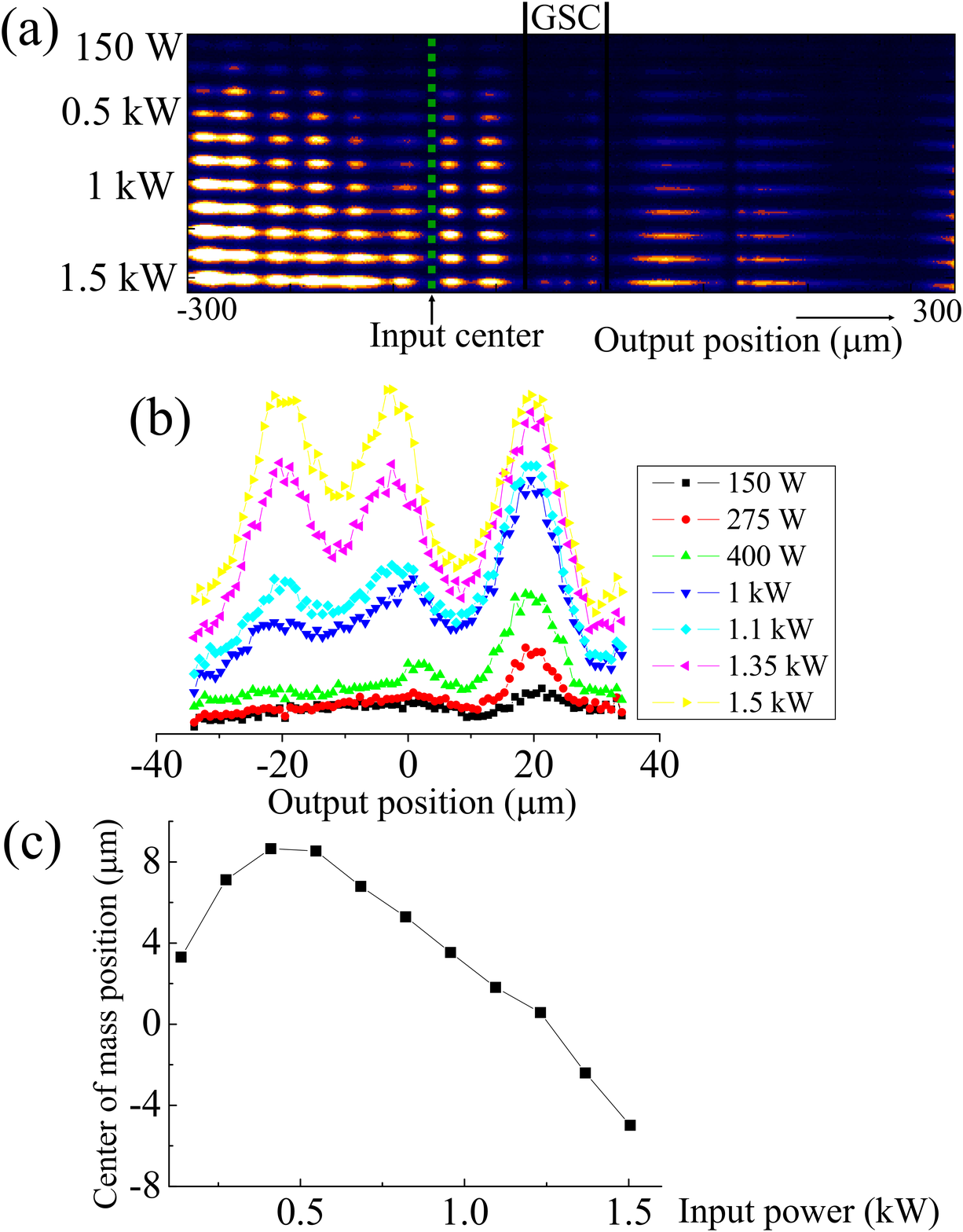}}
\caption{(Color online) Experimental results with $\theta$=10 mrad.
(a) Output facet images at different input powers. (b) Power
integrated along the vertical direction inside the GSC structure.
(c) Power distribution's center of mass position as a function of
the input power.}
\end{figure}

To realize the NLT switching we have fabricated 15 mm-long nonlinear
AlGaAs samples with a cross section as in Fig. 2(a). The excitation
was by 100 fs laser pulses, at a 1 kHz repetition rate, generated by
a Spectra Physics OPA, and yielding up to 10 kW peak power at a
wavelength of 1520 nm. The input beam was shaped to a narrow 20
$\mu$m-wide Gaussian at the input facet, and was coupled into the
sample's homogeneous region. The beam was tilted at various angles
$\theta$ relative to the input facet towards the GSC. Variable
neutral density filters were used to attenuate the input power, and
a Vidicon camera was used to image the sample's output facet.

Experimental results for $\theta$=10 mrad are presented in Fig. 4.
The output facet images at different input powers [Fig. 4(a)], and
the power distributions inside the structure, integrated along the
vertical direction [Fig. 4(b)], clearly show a change in the output
trapped power distribution as a function of the input power (while
for low input powers only reflection and transmission wave packets
are observed, in comparison to which the power inside the GSC is
negligible). The corresponding output power's center of mass
positions are shown in Fig. 4(c). The center of mass dynamics agree
with the simulation results of Fig. 3, and demonstrate an initial
shift of the center of mass towards the right site, in which most of
the power is trapped, followed by a shift towards the central and
left sites as the input power is further increased. The trapping and
switching dynamics are extremely sensitive to the input angle
$\theta$. Figure 5 shows the results of additional measurements, at
different input angles, with all other parameters as in Fig. 4. It
is evident from Figs. 4 and 5 that the extent of switching \emph{at
nonlinear power levels} (above 0.8 kW) changes as a function of the
input angle, while it is suppressed altogether at some input angles
(namely, below 5 mrad and above 20 mrad). In the latter case the
trapping is static. However, there is a broad range of excitation
angles (6-15 mrad) in which significant switching of the trapped
power is observed.

\begin{figure}[t]
\centerline{\includegraphics[width=8cm]{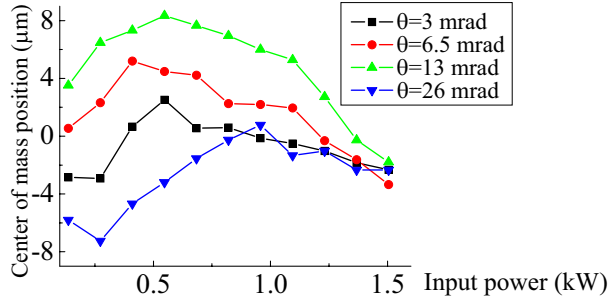}}
\caption{(Color online) The output center of mass position as a
function of the input power in experimental measurements with
several excitation input angles.}
\end{figure}

In conclusion, we have demonstrated power dependent switching of the
nonlinear trapping in 2D wave scattering by local multi-site guiding
photonic potentials. Our experimental results prove that AlGaAs
provides sufficient nonlinearity strengths for observable switching
of the nonlinearly trapped power. This observation of NLT
manipulation between GSCs, monolithically embedded in nonlinear
planar waveguides, with the excitation power as a single parameter,
can prove useful in optical switching applications. Our numerical
simulations, which agree with the experimental results, also imply
that tailored local potential profiles can lead to more prominent
switching characteristics, which can be valuable in such
applications.

This research was supported by the Israel Science Foundation through
grant numbers 8006/03 and 944/05, and by NSERC in Canada.

\end{document}